\documentclass[final,3p,times,twocolumn]{elsarticle}

\usepackage{siunitx}

\usepackage{amssymb, amsmath, amsfonts, amsthm}
\usepackage{mathtools}
\usepackage{bbm}

\usepackage{paralist} 

\usepackage{url}

\usepackage{braket}

\usepackage{algorithm}
\usepackage{algpseudocode}

\usepackage[usestackEOL]{stackengine}

\usepackage{subcaption}
\usepackage{tikz}
\usepackage{tikz-3dplot}
\usepackage{circuitikz}
\usepackage{pgfplots}
\usetikzlibrary{positioning, shapes, fit, calc, external, arrows, matrix}
\usetikzlibrary{tikzmark}
\pgfplotsset{compat=1.15}

\usepackage{ifthen}

\DeclareMathOperator*{\argmax}{arg\,max}

\newcommand{\todo}[1]{{\color{black}{#1}}}

\begin{document}

\begin{frontmatter}



\title{Hierarchical Bayesian approach for adaptive integration of Bragg peaks in time-of-flight neutron scattering data}

\affiliation[DAML]{organization={Data Analysis and Machine Learning Group, Oak Ridge National Laboratory},
            addressline={1 Bethel Valley Rd},
            city={Oak Ridge},
            postcode={37830},
            state={TN},
            country={USA}}

\affiliation[SNS]{organization={Single Crystal Diffraction, Oak Ridge National Laboratory},
            addressline={1 Bethel Valley Rd},
            city={Oak Ridge},
            postcode={37830},
            state={TN},
            country={USA}}

\affiliation[COMPEARTH]{organization={Computational Earth Sciences Group, Oak Ridge National Laboratory},
            addressline={1 Bethel Valley Rd},
            city={Oak Ridge},
            postcode={37830},
            state={TN},
            country={USA}}

\affiliation[AI]{organization={Analytics and AI Methods at Scale, Oak Ridge National Laboratory},
            addressline={1 Bethel Valley Rd},
            city={Oak Ridge},
            postcode={37830},
            state={TN},
            country={USA}}

\author[DAML]{Viktor Reshniak\corref{cor1}}
\author[SNS]{Xiaoping Wang}
\author[DAML]{Guannan Zhang}
\author[COMPEARTH]{Siyan Liu}
\author[AI]{Junqi Yin}

\cortext[cor1]{Corresponding author}


\begin{abstract}

The Spallation Neutron Source (SNS) at Oak Ridge National Laboratory (ORNL) operates in the event mode.
Time-of-flight (TOF) information about each detected neutron is collected separately and saved as a descriptive entry in a database enabling unprecedented accuracy of the collected experimental data.
Nevertheless, the common data processing pipeline still involves the binning of data to perform analysis and feature extraction. 
For weak reflections, improper binning leads to sparse histograms with low signal-to-noise ratios, rendering them uninformative.
In this study, we propose the Bayesian approach for the identification of Bragg peaks in TOF diffraction data.
The method is capable of adaptively handling the varying sampling rates found in different regions of the reciprocal space. Unlike histogram fitting methods, our approach focuses on estimating the true neutron flux function. 
We accomplish this by employing a profile fitting algorithm based on the event-level likelihood, along with a multiresolution histogram-level prior. 
By using this approach, we ensure that there is no loss of information due to data reduction in strong reflections and that the search space is appropriately restricted for weak reflections.
To demonstrate the effectiveness of our proposed model, we apply it to real experimental data collected at the TOPAZ single crystal diffractometer at SNS.
\end{abstract}

\begin{keyword}
Bragg peaks \sep peak integration \sep neutron scattering \sep time of flight diffraction \sep Bayesian



\end{keyword}

\end{frontmatter}


\section{Introduction and background}

\paragraph{Spallation Neutron Source}
The Spallation Neutron Source (SNS) at Oak Ridge National Laboratory (ORNL) is presently the most powerful accelerator-driven neutron source in the world capable of producing a \todo{$1.5$~MW, $1$~GeV} beam of protons delivered in $\sim 1\mu s$ long pulses to a liquid mercury neutron-producing spallation target \cite{MASON2006955,HENDERSON2014610}.
The spalled neutrons are slowed down in a moderator and guided through beams to state-of-the-art instruments, which offer a diverse range of capabilities to researchers across multiple scientific disciplines \cite{sns}. Operating since 2006, the SNS currently houses 18 instruments for its user program, providing a wide array of neutron scattering techniques that have facilitated significant discoveries and innovations in areas such as energy, environment, health, and technology. Additionally, ORNL has plans to construct a Second Target Station at the SNS, which will enhance neutron capabilities and address emerging scientific challenges.

\begin{figure}[tb]
    \centering
    \tikzsetnextfilename{pynn_scatt}
    \begin{tikzpicture}[trim axis left,trim axis right]
	    \tikzstyle{every node}=[font=\scriptsize]
 		\begin{axis} [
 			width=1.2\linewidth,
 			height=1\linewidth,
 			ymode = normal,
 			xmin = -0.2,
 			xmax = 1,
 			ymin = 0,
 			ymax = 1,
 			axis x line = none,
 			axis y line = none,
 			axis line style={draw=none},
 			tick style={draw=none},
 			]
 			\fill [gray!20] (0, 0) rectangle (1, 1);
 			\draw [very thick] (0, 0) -- (0, 1);
 			\node [left, above, rotate=90] at (-0.005, 0.75) {Source};
 			
 			\def\slope{0.3}
 			\def\rad{0.06}
 			
            \foreach \i in {0,...,10}{
                \pgfmathsetmacro{\x}{\rad+\i*0.14}
                \foreach \j in {0,...,10}{
                    \pgfmathsetmacro{\y}{\j*0.14}
                    \ifodd\i
                        \pgfmathsetmacro{\yy}{\y+\rad}
                        \pgfmathsetmacro{\r}{rand}
                        \pgfmathsetmacro{\a}{\x+\rad*(cos(\r*360))}
                        \pgfmathsetmacro{\b}{\yy+\rad*(sin(\r*360))}
                        \edef\tempa{\noexpand\draw (\x,\yy) circle (\rad);}\tempa
                        \edef\tempb{\noexpand\draw[black,fill=black] (\x,\yy) circle (0.007);}\tempb
                        \edef\tempc{\noexpand\draw[black,fill=black] (\a,\b) circle (0.005);}\tempc
                    \else
                        \pgfmathsetmacro{\r}{rand}
                        \pgfmathsetmacro{\a}{\x+\rad*(cos(\r*360))}
                        \pgfmathsetmacro{\b}{\y+\rad*(sin(\r*360))}
                        \edef\tempa{\noexpand\draw (\x,\y) circle (\rad);}\tempa
                        \edef\tempb{\noexpand\draw[black,fill=black] (\x,\y) circle (0.007);}\tempb
                        \edef\tempc{\noexpand\draw[black,fill=black] (\a,\b) circle (0.005);}\tempc
                    \fi
                }
            }
            \def\xa{0.14}
            \pgfmathsetmacro{\ya}{\slope*(\xa+0.2)}
            \edef\tempa{\noexpand\draw[black!70, very thick] (-0.2,0.0) -- (\xa,\ya) node[currarrow, color={black!70}, pos=0.25, sloped]{} node[pos=0.25, sloped, above]{Electron};}\tempa
            \edef\tempb{\noexpand\draw[black!70, very thick] (0.14,\ya) -- (0.27,0.01) node[currarrow, color={black!70}, pos=1.0, sloped]{};}\tempb
            
            \def\xa{0.35}
            \def\yi{0.16}
            \pgfmathsetmacro{\ya}{\yi+\slope*(\xa+0.2)}
            \edef\tempa{\noexpand\draw[black, very thick] (-0.2,\yi) -- (\xa,\ya)  node[currarrow, color={black}, pos=0.15, sloped]{} node[pos=0.15, sloped, above]{X-ray};}\tempa
            \edef\tempb{\noexpand\draw[black, very thick] (\xa,\ya) -- (0.75,0.03) node[currarrow, color={black}, pos=1.0, sloped]{};}\tempb
            
            \def\xa{0.68}
            \def\yi{0.35}
            \pgfmathsetmacro{\ya}{\yi+\slope*(\xa+0.2)}
            \edef\tempa{\noexpand\draw[blue!80!black, very thick] (-0.2,\yi) -- (\xa,\ya) node[currarrow, color={blue!80!black}, pos=0.08, sloped]{} node[pos=0.08, sloped, above]{Neutron};}\tempa
            \edef\tempb{\noexpand\draw[blue!80!black, very thick] (\xa,\ya) -- (0.97,0.85) node[currarrow, color={blue!80!black}, pos=1.00, sloped]{};}\tempb
            
            \def\xa{\rad+3*0.14}
            \def\ya{\rad+5*0.14}
            \pgfmathsetmacro{\yi}{\ya-\slope*(\xa+0.2)}
            \edef\tempa{\noexpand\draw[blue!80!black, very thick] (-0.2,\yi) -- (\xa,\ya) node[currarrow, color={blue!80!black}, pos=0.1, sloped]{} node[pos=0.1, sloped, above]{Neutron};}\tempa
            \edef\tempb{\noexpand\draw[blue!80!black, very thick] (\xa,\ya) -- (0.54,0.99) node[currarrow, color={blue!80!black}, pos=1.00, sloped]{};}\tempb
 		\end{axis}
 	\end{tikzpicture}
    \caption{Schematic illustration of different mechanisms of probe beam interactions with materials \cite{pynn1990neutron,sivia2011elementary}.}
    \label{fig:pynn_scatt}
\end{figure}
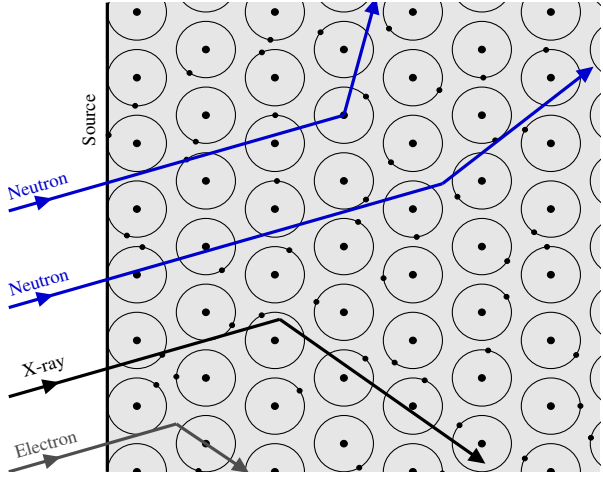

In a typical scattering experiment, a probe beam is directed at a material sample, triggering a specific interaction mechanism with its atoms, see Figure~\ref{fig:pynn_scatt}.
The outcome of this interaction is observed at a detector as the interference pattern 
of particle wavefunctions scattered from different nuclei.
The benefit of neutrons as scattering particles is in their electrical neutrality that allows them to penetrate deep into atoms and interact with nuclei via the strong nuclear force \cite{sivia2011elementary}.
Neutrons are also sensitive to light atoms and can distinguish between similar-mass elements within a composite material \cite{carpenter_loong_2015}.
On the downside, the brightness of the neutron sources is commonly orders of magnitude lower then the brightness of X-ray beams, leading to relatively weak reflections.
This possesses certain challenges for the collection, processing, and analysis of neutron scattering data.

\paragraph{Basic theory of thermal neutron scattering}

For slow neutrons ($\leq 0.025$ eV), the interaction with scattering potential is weak and short ranged.
This ensures the validity of Born approximation which assumes that both the incident and scattered waves are plane waves.
Such idealization enables relatively simple geometric interpretation of interference patterns utilizing the principle of superposition of plane waves.
Particularly, the general form of the partial differential cross section $d^2\sigma/d\Omega dE_f$, which indicates the probability of observing a scattered neutron at a solid angle $\Omega$ and at energy $E_f$, is expressed in terms of the spatio-temporal Fourier transforms as follows \cite{boothroyd2020principles}
\begin{align*}
    \frac{d^2\sigma(\mathbf{Q},\omega)}{d\Omega dE_f}
    &\simeq \frac{|\mathbf{k}_f|}{|\mathbf{k}_i|} S(\mathbf{Q},\omega)
    = \frac{|\mathbf{k}_f|}{|\mathbf{k}_i|} \int_{-\infty}^{\infty} {I(\mathbf{Q},t)} e^{-i\omega t} dt,
    \\[1em]
    I(\mathbf{Q},t) &= 
    \big\langle \hat{A}^\dag_0 \hat{A}_t \big\rangle,
    \quad
    \hat{A}_t = \braket{\sigma_f|\hat{V}(\mathbf{Q},t)|\sigma_i},
\end{align*}
where $\mathbf{Q}=\mathbf{k}_i-\mathbf{k}_f$ is the scattering vector, 
$\hbar\omega=E_i - E_f$ is the energy transfer, 
and $\sigma_i/\sigma_f$ are the initial/final polarization states of the scattered neutron.
$V(\mathbf{Q})$ is the spatial Fourier transform of the scattering potential
\begin{align*}
    V(\mathbf{Q}) &= \int_{\mathbb{R}^3} V(\mathbf{r}) e^{i\mathbf{Q}\cdot \mathbf{r}} d\mathbf{r},
\end{align*}
and $\hat{V}(\mathbf{Q},t)$ is a time-dependent Heisenberg operator describing temporal evolution of an observable $V(\mathbf{Q})$.
The correlation function $I(\mathbf{Q},t)$ is called the intermediate scattering function, it measures how the values of a physical quantity $\hat{A}_t$ relate at different time instances.
The scattering function $S(\mathbf{Q},\omega)$ is a time Fourier transform of $I(\mathbf{Q},t)$ measuring the spectrum of spontaneous fluctuations of $\hat{A}_t$.
The energy resolved output of a scattering experiment is thus proportional to the neutron-related factor $|\mathbf{k}_f|/|\mathbf{k}_i|$ and the scattering function $S(\mathbf{Q},\omega)$ which represents the physical properties of the sample.
In total scattering experiments, the measured quantity is the structure factor $$S(\mathbf{Q}):=\int S(\mathbf{Q},\omega) d\omega = \langle|\hat{A}|^2\rangle = |\langle\hat{A}\rangle|^2 + \tilde{S}(\mathbf{Q})$$ which accounts for both static and dynamic instantaneous correlations.

The ultimate goal of a scattering experiment is to determine the scattering potential $V(\mathbf{r})$ of the target sample from the measured cross-sections. 
Accomplishing this task in its general form is extremely challenging and further simplifications are required to make it amenable for interpretation and analysis.
In the case of thermal neutrons and crystalline samples,  
the scattering is close to elastic ($|\mathbf{k}_i|\approx|\mathbf{k}_f|$, $\hbar\omega\approx 0$) validating the static approximation $S(\mathbf{Q})\approx|\langle\hat{A}\rangle|^2$ which disregards the dynamic correlations.
For periodic rigid structures, the Fermi pseudopotential is an adequate simplification of the interaction potential
\begin{align*}
    &V(\mathbf{r}) \simeq \sum_{j} b_j \delta(\mathbf{r}-\mathbf{r}_{j}),
\end{align*}
where $b_j$ are the scattering lengths of the nuclei. 
The coherent differential cross-section $(d\sigma/d\Omega)_{coh}$ of elastic scattering from such potential is given by the structure factor
\begin{align*}
    S(\mathbf{Q})
    = \left| \sum_j b_j \exp(i\mathbf{Q}\cdot\mathbf{r}_j) \right|^2 
\end{align*}
that is negligible for all $\mathbf{Q}$ except those which coincide with the reciprocal lattice vectors $\mathbf{H}$ satisfying $\exp(i\mathbf{H}\cdot\mathbf{r}_j)=1$ for all $\mathbf{r}_j$.
Therefore, the strong interference occurs for $\mathbf{Q}=\mathbf{H}$, known as the Laue condition, producing a perfect interference pattern of sharp peaks in $\mathbf{Q}$-space.
These peaks are of primary practical interest because they contain important structural information of the atomic arrangement in a sample. 
Unfortunately, the inevitable traces of dynamic self-correlations in a sample are the source of incoherent scattering which produces isotropic background devoid of structural information.
Various other factors such as structural disorder, short-range order, point defects, multiple scattering, thermal fluctuations, or limited equipment resolution contribute to peak broadening, appearance of secondary peaks, diffuse scattering, and other background artifacts as well.
The correct identification and separation of structurally significant features from meaningless background artifacts constitut the primary challenge in experimental data processing.
Subsequent tasks involve peak integration to determine structure factors and structure refinement for ascertaining atomic arrangements.

\paragraph{TOPAZ instrument}

\begin{figure*}[tb]
    \centering
    \includegraphics[width=0.9\textwidth]{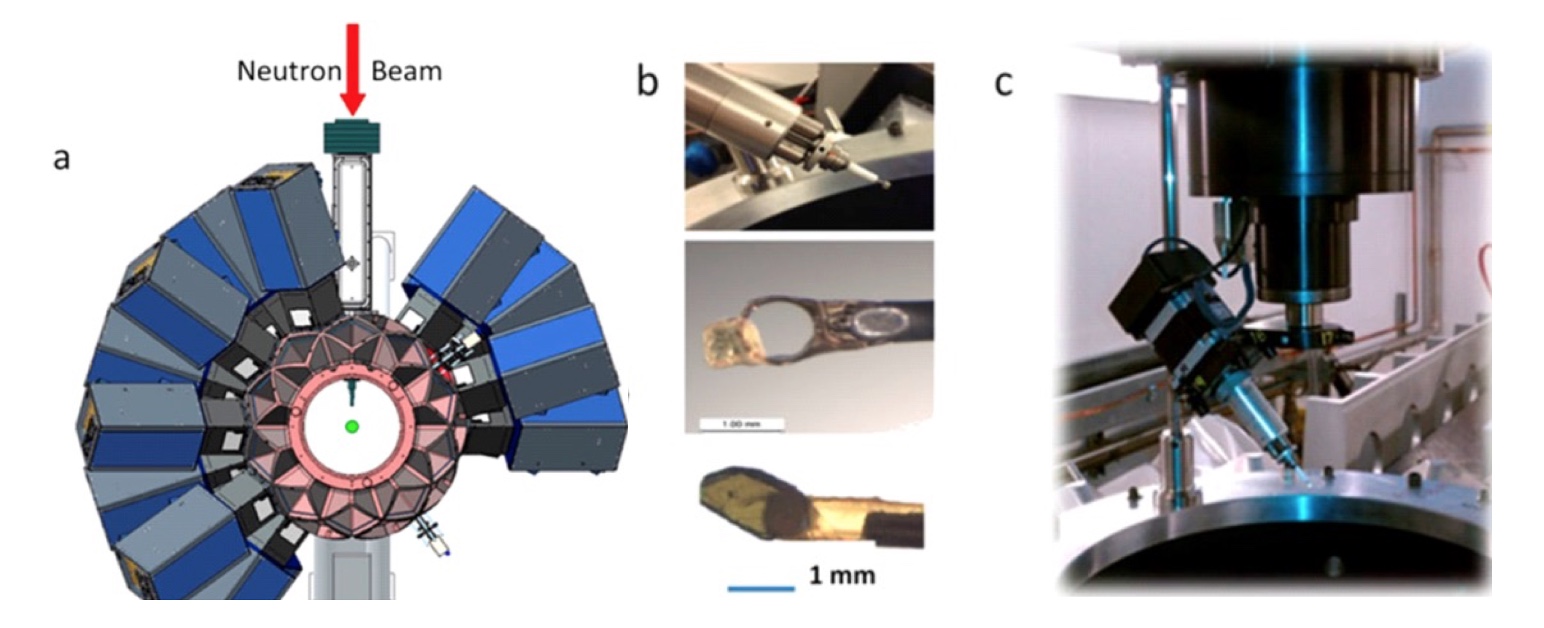}
    \caption{Top view of the TOPAZ layout with area detectors populated on 25 of 48 detector ports on the detector array tank (DAT); (b) single crystals mounted on various sample pins; (c) TOPAZ ambient goniometer with fixed chi and unrestricted 360° rotations on phi and omega axes. The sample is lowered into the neutron beam position at the center of the DAT for data collection.}
    \label{fig:TOPAZ}
\end{figure*}

The TOPAZ instrument at SNS is a single-crystal TOF Laue diffractometer \cite{topaz}. 
It has 25 Anger camera detectors, each with active areas of $15\times15$~cm and $256\times256$~pixels, arranged on a spherical detector array tank, see Figure~\ref{fig:TOPAZ}. 
As a diffractometer, TOPAZ is designed to measure predominantly elastic scattering with inelastic component minimized by careful configuration of experiments, e.g., by using low temperatures.

In the Laue method, the polychromatic neutron beam is diffracted from a crystal that is fixed relative to $\mathbf{k}_i$.
The Bragg peaks are detected for all crystallographic planes and those neutron wavelengths that satisfy the Bragg's law for constructive interference
\begin{align*}
    n\lambda=2d_{hkl}\sin\theta,
\end{align*}
where $n$ is a positive integer, $\lambda$ is the neutron wavelength, $d_{hkl}$ is the crystal interplanar spacing, and $\theta$ is the incident angle of a set of crystlal lattice planes a distance $d_{hkl}$.
The time-of-flight technique allows separation of the neutron events with different wavelengths by means of the de Broglie relationship
\begin{align*}
    \lambda 
    = \frac{ht}{m(L_1+L_2)},
\end{align*}
where $h$ is a Plank constant, $m$ is the mass of neutron, $t$ is the time of flight, and $L_1+L_2$ is the total path from the source to detector.
Essentially this allows identification of reflections from specific crystallographic planes.
Additionally, the event mode of data recording enables unprecedented flexibility for the analysis of the collected data since the reduction is performed post factum ensuring no loss of information.

The data analysis tool used by instruments at SNS is \textit{Mantid}, the open-source software framework created to manipulate and analyse neutron scattering and muon spectroscopy data \cite{ARNOLD2014156}.
A common data reduction workflow for single crystal TOF measurements using \textit{Mantid} is described, for instance, in \cite{Savici:fs5205}.

\paragraph{Peak indexing}
The Laue condition for diffraction in laboratory coordinates reads as
\begin{align*}
    \mathbf{Q}_{lab} := \mathbf{k}_f-\mathbf{k}_i = 2\pi \mathsf{RUB} 
    \begin{pmatrix}
        h\\
        k\\
        l
    \end{pmatrix},
\end{align*}
where $\mathsf{R}$ is a goniometer rotation matrix, $\mathsf{U}$ is a rotation matrix that determines the crystal orientation in the instrument's goniometer-head-fixed orthonormal coordinate system in the laboratory frame with all goniometer angles equal to zero, and $\mathsf{B}$ is the reciprocal basis matrix which transforms a reciprocal lattice vector $\mathbf{H}$ into an orthonormal Cartesian coordinate system
\begin{align*}
    \mathbf{H} 
    := h\mathbf{a}^* + k\mathbf{b}^* + l\mathbf{c}^* 
    = \mathsf{B} 
    \begin{pmatrix}
        h\\
        k\\
        l
    \end{pmatrix},
\end{align*}
where $hkl$ are the Miller indices of crystallographic planes.

The $\mathsf{UB}$ matrix depends on the crystal structure parameters and positioning of the crystal inside goniometer.
Once it is known, the crystal can be positioned to satisfy Bragg condition for any desired $hkl$ plane.
Conversely, for the known crystal orientation, $\mathsf{UB}$ matrix is used to determine the expected position of a Bragg reflection in $hkl$ space.
The \verb|PredictPeaks| algorithm in \textit{Mantid} operates by calculating the scattering direction for a particular $hkl$ (given the $\mathsf{UB}$ matrix) and determines whether that hits a detector.
The $hkl$ values to try are determined from the range of acceptable $d$-spacings.

\paragraph{Peak integration}

Predicted peaks can be integrated using various methods.
Theoretical peak shape identification is possible in some special cases \cite{CooperNathans1968}, but such calculations are too cumbersome due to large number of crystal and experimental parameters that determine the shape of a peak.
Practical methods involve a range of empirical techniques.
The simplest approach is to assigns pixels at the perimeter of the integration box to the background and the remaining pixels to the peak region.
However, this method provides no feedback and relies on manual selection of the peak region.
In the `dynamic mask' technique of \cite{Sjolin:a19718}, a peak region is assigned to the detector pixels with intensities above the background noise determined separately.
This method allows peak regions of arbitrary shapes but is otherwise similar to the classical $\sigma(I)/I$ criterion.
The latter is based on the observation that the relative variance of the corrected peak intensity attains its minimum near the true boundary of the peak region \cite{Lehmann:a10706}. 
The method was originally proposed for one-dimensional profiles and then extended to the integration of multi-dimensional peaks of given shape \cite{Wilkinson:sr0074}.
Both approaches require relatively strong reflections rendering them inefficient for TOF neutron data.
Seed-skewness method in \cite{Bolotovsky1995,Peters:hx5002} employs a different statistical criterion based on the observation of higher skewness of pixel intensities in peak region compared to the skewness of the noise distribution.
The method starts with an initial seed of the peak region and keeps adding pixels until the skewness outside the peak region reaches minimum.
It was shown to be superior to standard-box and dynamic mask methods for weak reflections, but tends to produce small masks for weaker peaks.

The methods above have been proposed for histogram data collected on instruments with limited resolution and assume the predefined integration box containing a single peak with background free from artifacts.
Instead, peak integration routines in \textit{Mantid} operate directly on the event data in $\mathbf{Q}$-space \cite{Schultz:he5647}.
The peaks are estimated either by calculating ellipsoids of inertia, see e.g. \cite{Spencer1980}, or by fitting a three-dimensional profile such as Ikeda-Carpenter function with a bivariate Gaussian \cite{Sullivan:mn5117}.
The first approach is very fast, the second is more accurate but requires careful supervision for parameter initialization.
Both approaches rely on the user feedback and continue to struggle with weak reflections and background artifacts.
Data driven neural network and ML-based approaches have been also proposed \cite{Sullivan:fs5173,Liu:fs5198}.

In this work, we aim to develop an automated routine for peak identification and integration that 
\begin{inparaenum}[1)]
    \item minimizes amount of required supervision,
    \item is consistent for both strong and weak reflections,
    \item is adaptive to the size and shape of different peaks,
    \item is capable to handle background artifacts, and
    \item does not require pre-trained model or data labeling.
\end{inparaenum}

The rest of the paper is organized as follows.
In section~\ref{sec:model}, we describe the proposed mathematical model.
Section~\ref{sec:alg} provides implementation details.
Section~\ref{sec:res} concludes with results.

\section{Mathematical model}
\label{sec:model}

Denote by $\lambda(q,\tau)$ the true flux of neutrons at a time instance $\tau$ and at a point $q := hkl$  in the reciprocal space $\mathbf{Q}$. 
The rate of neutron events in any region $\mathbf{Q}_i \subset \mathbf{Q}$ hence can be expressed as
\begin{align*}
    \Lambda_i(\tau) = \int_{\mathbf{Q}_i} \lambda(q,\tau) dq,
\end{align*}
and the total number of neutron events in $\mathbf{Q}_i \subset \mathbf{Q}$ detected up to time $T$ is given by
\begin{align}\label{eq:flux}
    \Lambda_i = \int_0^T\int_{\mathbf{Q}_i} \lambda(q,\tau) dq d\tau.
\end{align}

Given the collection of $N_e$ detected neutron events $\mathcal{D}=\{d_i:=(q_i,\tau_i),i=1,\hdots,N_e\}$, our objective is to determine the flux function $\lambda(q,\tau)$ that is most likely to generate the observed outcome. 
Applying the Bayesian formulation \cite{gelman2013bayesian}, the problem can be solved by finding the maximum a posteriori (MAP) estimate of $\lambda(q,\tau)$, which involves
\begin{align}\label{eq:MAP}
    \lambda^*_{MAP} := \argmax_{\lambda} p\big(\lambda|\mathcal{D}\big) = \argmax_{\lambda} \log p\big(\lambda|\mathcal{D}\big),
\end{align}
where, according to the Bayes' rule, one has
\begin{align*}
    p\big(\lambda|\mathcal{D}\big) = \frac{\mathcal{L}\big(\mathcal{D}|\lambda\big)p\big(\lambda\big)}{\int\mathcal{L}\big(\mathcal{D}|\lambda\big)p\big(\lambda\big)d\lambda}
\end{align*}
and hence
\begin{align*}
    \log p\big(\lambda|\mathcal{D}\big) \simeq \log\mathcal{L}\big(\mathcal{D}|\lambda\big) + \log p\big(\lambda\big).
\end{align*}
Here, $\mathcal{L}\big(\mathcal{D}|\lambda\big)$ is the likelihood function that encodes the statistical evidence of observing data $\mathcal{D}$ generated by given $\lambda$, and $p\big(\lambda\big)$ represents the prior that encodes the modeling assumptions or prior beliefs for the flux function.

A closely related formulation involves maximizing the following objective 
\begin{align}\label{eq:loss_reg}
    \max_{\lambda} L\big(\mathcal{D},\lambda\big) + \alpha \cdot R\big(\lambda\big), \quad \alpha>0.
\end{align}
In this case, the objective function consists of the data fitting term $L\big(\mathcal{D},\lambda\big)$ and a regularization term $R(\lambda)$ that promotes the desired structure of $\lambda$. 
The choice of one formulation over another is largely motivated by the adopted interpretation of the data generating process and model constraints.
For example, the common regularized least-squares problem $\min_{\lambda}\sum_i\|y_i-f(x_i,\lambda)\|^2+\alpha\|\lambda\|^2$ is equivalent to the Gaussian measurement error model $y_i-f(x_i,\lambda)\sim\mathcal{N}(0,\sigma I)$ with a Gaussian prior on parameters $\lambda\sim\mathcal{N}(0,\sigma\alpha^{-1} I)$.
This equivalence holds for various scenarios, such as the utilization of more general Gaussian process priors or sparsity-inducing priors \cite{seeger2004gaussian, figueiredo2001adaptive}.

The impact of a prior is asymptotically diminishing in the limit of large data sample size that leads to a more concentrated likelihood. 
The analogy with regularization in \eqref{eq:loss_reg} shows that the corresponding contributions of the data and prior to the objective function depend on the relative uncertainty $\sigma\alpha^{-1}$ of the two distributions.
For instance, the uniform prior $p\big(\lambda\big)=const$ does not favor any particular realization of $\lambda$ and corresponds to the choice of $\alpha=0$ in \eqref{eq:loss_reg}.
In both cases, the prior does not contribute to the MAP estimate in \eqref{eq:MAP}
and the inference is solely determined by the data through the likelihood function. 
In this case, we seek the maximum likelihood estimate (MLE) of $\lambda$
\begin{align}\label{eq:MLE}
    \lambda^*_{MLE} := \argmax_{\lambda} \log\mathcal{L}\big(\mathcal{D}|\lambda\big).
\end{align}

MLE is the commonly used method for fitting profiles to data based on counting statistics from binned detectors. 
In this setting, the data is given in the form of a histogram
$\mathcal{H}=\{n_i,i=1,\hdots,N_b\}$, where $n_i$ represents the number of events observed in the $i$-th bin with 
$\sum_{i=1}^{N_b}n_i=N_e$. 
The joint probability of bin counts can be modelled with a multinomial distribution $p(n_i)\sim \prod_{i=1}^{N_b} p_i^{n_i}$ such that $\sum_{i=1}^{N_b}p_i=1$ and $\Lambda_i=N_ep_i$. Directly applying the MLE approach to the multinomial distribution gives the objective function
\begin{align*}
    \text{Multinomial:}\qquad\log\mathcal{L}\big(\mathcal{H}|\Lambda_i\big) \sim \sum_{i=1}^{N_b} n_i \log\Lambda_i.
\end{align*}
Alternatively, assuming many bins with $p_i\ll 1$, the correlations between bins are negligible, and 
each $n_i$ can be well-approximated by a Poisson distribution $\mathcal{P}(n_i)$, i.e., $p(n_i)\sim \Lambda_i^{n_i}\exp(-\Lambda_i)$, where again $\Lambda_i=N_ep_i$ is the average number of counts in the detector $i$. This results in the following objective function
\begin{align}\label{eq:poisson_loglikelihood}
    \text{Poisson:}\qquad\log\mathcal{L}\big(\mathcal{H}|\Lambda_i\big) \sim \sum_{i=1}^{N_b} ( n_i \log\Lambda_i - \Lambda_i ).
\end{align}
In the limit of large $n_i$, one has $\sigma^2(n_i)\approx \Lambda_i \sim n_i$ and the normality assumption is valid leading to $\mathcal{P}(n_i)\approx\mathcal{N}(n_i,n_i)$ and well-known $\chi^2$ goodness-of-fit tests
\begin{align*}
    \text{Pearson }\chi^2:\qquad&\log\mathcal{L}\big(\mathcal{H}|\Lambda_i\big) \sim \sum_{i=1}^{N_b} \frac{(n_i-\Lambda_i)^2}{\Lambda_i^2},
    \\
    \text{Neumann }\chi^2:\qquad&\log\mathcal{L}\big(\mathcal{H}|\Lambda_i\big) \sim \sum_{i=1}^{N_b} \frac{(n_i-\Lambda_i)^2}{n_i^2}.
\end{align*}

All four likelihood models are consistent and asymptotically equivalent \cite{eadie1971statistical}. 
However, the Gaussian approximation, even though commonly employed in practice, yields skewed and biased fits both within and outside the Poisson regime. 
These challenges are particularly noticeable in the low-count limit, where the assigned uncertainty to the data point becomes unphysically low, biasing the model towards zero and allowing for high probabilities of negative values.
On the other hand, the Poisson and Multinomial models have been found to be less stable than Gaussian model when the fitting function does not fully capture the observed data \cite{lass2021multinomial}. 
These issues become even more pronounced in multiple dimensions, as successful fitting requires a larger amount of data to accurately represent the features of interest.

Multidetector time-of-flight measurements, such as those conducted at the TOPAZ instrument, provide the capability to explore a wide range of reciprocal space simultaneously. 
However, these measurements often result in an uneven distribution of neutron event density. 
Analyzing regions with low event counts poses a considerable challenge due to the limited amount of information available in those areas. 
Nonetheless, the underlying physics of the process suggests the existence of some level of spatio-temporal correlation in the measured data.
\todo{Figure ?} clearly demonstrates limitations of the purely data-driven approach for the correct identification of an isolated Gaussian peak.
It underscores the significance of incorporating prior knowledge into the model, particularly in situations with limited data.
Taking this into consideration, the ultimate desired goal of the data processing step is to enable exploitation of the inherent data correlations, resulting in a single model capable to fill the gaps and consistently represent the collected multiresolution data.
The rest of this section presents an attempt towards achieving this goal.

\subsection{Data model and likelihood}

Neutron scattering measurements involve counting events in the detector region over a specific time interval. 
Historically, due to the finite resolution of detectors, the data was collected as a histogram with fixed-size bins corresponding to separate detectors or detector pixels.
The widely accepted assumption of stochastic independence among individual events justifies the adoption of the Poisson error model for the counts in such histogram bins \cite{lass2021multinomial,boothroyd2020principles}.
The fitting of the selected model to the data was then carried out using one of the MLE approaches described in the previous section.

The ever increasing accuracy of modern detectors, including TOF measurements performed at SNS, unlocks new possibilities for interpreting high-resolution features extracted from experimental data.
However, in practice, the common processing pipeline still involves the binning of data to perform analysis and feature extraction \cite{ARNOLD2014156}.
One advantage of this approach is that the resolution of the binned data can be made arbitrarily high. However, a drawback is the inevitable loss of information from the original event data, particularly for strong reflections. 
Furthermore, when dealing with weak reflections, improper binning leads to very sparse histograms with low signal-to-noise ratios, rendering them uninformative.

Instead of relying on the particular binned representation, we interpret the true data generating process as a stochastic point process \cite{daley2003introduction, snyder2012random}.
Specifically, we consider the spatially inhomogeneous Poisson process with a rate function $\lambda(q,\tau)$ defined in \eqref{eq:flux}.
The general form of the log likelihood function for this event-level process is given by
\begin{align*}
    \log\mathcal{L}(\mathcal{D}|\lambda) \sim \sum_{i=1}^{N_e} \log \lambda(q_i,\tau_i) - \int_0^T \int_{\mathbf{Q}} \lambda(q,\tau) dq d\tau.
\end{align*}

The likelihood over an arbitrary binned detector for data collected up to time $T$ can then be calculated as
\begin{align}\label{eq:binned_MLE}
    \log\mathcal{L}(\mathcal{H}_s|\lambda) \sim \sum_{i=1}^{N_b}  n_i \log \lambda_i - \int_{\mathbf{Q}} \lambda(q) dq,
\end{align}
where $\mathcal{H}_s$ denotes the histogram at scale $s$, $n_i$ is the number of events in the $i$-th bin of the histogram, $\lambda_i$ is the value of the flux at the center of the bin, and we assumed the time-homogeneous rate $\lambda(q,\tau):=\lambda(q)\tau$.

The likelihoods in \eqref{eq:binned_MLE} provide a one-parameter family of hierarchical data representations at different resolutions with $\mathcal{H}_L$ being the finest resolution with $\mathcal{H}_\infty:=\mathcal{D}$. 
The construction of the hierarchy can be also accompanied with an appropriate smoothing operation $\mathcal{H}_s:=G_s\ast\mathcal{H}_L$ leading to the proper scale-space representation of the data \todo{\cite{Lindeberg1990}}.

\subsection{Hierarchical prior}

The purpose of incorporating a prior in our construction is to utilize the available data to ``fill the gaps" in regions with low event counts to avoid ambiguity in the peak identification and to facilitate the overall optimization procedure.
One should not understand ``filling the gaps" literally, of course.
Instead, we need a prior for the model parameters to provide a robust initial guess with uncertainty levels that provide an adequate restriction on the search space without conflicting with the true collected data.
This task can be achieved in multiple ways.
If an oracle guess for the model parameters is available, one can simply restrict the range of admissible parameter values leading to the MLE formulation with parameter constraints.
If such an oracle is not available, the appropriate initialization should be estimated from the data itself.
For example, it can be provided by an independent model learned on the corpus of the previously collected and labeled data 
\todo{\cite{Sullivan:fs5173,Liu:fs5198}}.
Both approaches typically require a significant level of human involvement.

Alternatively, one might focus exclusively on the collected data and consider the problem
\begin{align}\label{eq:selfsim_MLE}
    \max_{\lambda} \log\mathcal{L}\big(\mathcal{D}|\lambda\big) + \alpha \cdot \int_{Q} d\big[\mathcal{H}_s, \lambda\big](q) dq,
\end{align}
where $d\big[\mathcal{H}_s, \lambda\big]$ is an appropriate non-local self-similarity feature metric \todo{\cite{reshniak2020}}.
This regularizer can be viewed as a form of a Gaussian prior with a non-standard distance function.
The clear benefit of this approach lies in its self-consistency and adaptivity.

In the current effort, we explore a simpler variant of the equation \eqref{eq:selfsim_MLE} by employing a multiresolution metric specific to each peak of interest. 
As before, we define $\mathcal{H}_L$ to be the histogram with the finest resolution
and introduce a hierarchy of histograms with increasing resolutions, denoted as $\mathcal{H}_s$ for $s=0,\hdots,L$, where $\mathcal{H}_0$ represents the coarsest histogram. 
This concept bears resemblance to multiresolution approaches in image processing, where image pyramids are utilized to extract scale-specific features \todo{\cite{Lindeberg1990}}. 
An analogy can also be drawn with multigrid solvers that aim to address errors at different scales individually \cite{DAROCHAAMARAL2004654,Briggs2000}.

The next step is to consider the 
posterior probability $p(\lambda|\mathcal{H}_0,\mathcal{H}_1,\hdots,\mathcal{H}_L)$ and apply the chain rule to get
\begin{align*}
    &p(\lambda|\mathcal{H}_0,\mathcal{H}_1,\hdots,\mathcal{H}_L)
    = \frac{p(\lambda,\mathcal{H}_0,\hdots,\mathcal{H}_L)}{p(\mathcal{H}_0,\hdots,\mathcal{H}_L)}
    \\
    &\sim\mathcal{L}(\mathcal{H}_L|\lambda,\mathcal{H}_0,\hdots,\mathcal{H}_{L-1}) p(\lambda,\mathcal{H}_0,\hdots,\mathcal{H}_{L-1}) 
    \\
    &= p(\lambda) \prod_{i=0}^{L} \mathcal{L}(\mathcal{H}_{i}|\lambda,\mathcal{H}_{0},\hdots,\mathcal{H}_{i-1})
\end{align*}
where $p(\lambda)$ is the prior encoding any available information regarding the flux.
Hence, the log posterior of the observed data at any resolution $s\geq0$ can be decomposed into 
\begin{align*}
    &\log p(\lambda|\mathcal{H}_0,\hdots,\mathcal{H}_s)
    \\
    &\sim \sum_{i=0}^s \log\mathcal{L}(\mathcal{H}_{i}|\lambda,\mathcal{H}_0,\hdots,\mathcal{H}_{i-1}) + \log p(\lambda)
    \\
    &= \log\mathcal{L}(\mathcal{H}_s|\lambda,\mathcal{H}_0,\hdots,\mathcal{H}_{s-1}) + \log p(\lambda|\mathcal{H}_0,\hdots,\mathcal{H}_{s-1}).
\end{align*}
Written this way, it allows to consider the posterior at scale $s-1$ as the prior at scale $s$.
We then use the following simplifying assumption for the conditional log-likelihood
\begin{align*}
    &\log p(\lambda|\mathcal{H}_0,\hdots,\mathcal{H}_s)
    \\
    &= \log\mathcal{L}(\mathcal{H}_s|\lambda) + \alpha\cdot\log p(\lambda|\mathcal{H}_0,\hdots,\mathcal{H}_{s-1}),
    \quad\alpha\geq 1,
\end{align*}
where we accounted for the higher certainty of coarser scale details by moving it from the conditional likelihood to the prior.
This recurrence expands to 
\begin{align*}
    &\log p(\lambda|\mathcal{H}_0,\hdots,\mathcal{H}_s)
    \\
    &= \log\mathcal{L}(\mathcal{H}_s|\lambda) + \sum_{i=0}^{s-1} \alpha^{s-i} \log\mathcal{L}(\mathcal{H}_i|\lambda) + \log p(\lambda),
\end{align*}
which means that likelihoods of coarser histograms are assigned larger weights accounting for the higher certainty of observing the neutron event in a larger detector region.

\begin{figure*}[tb]
    \centering
    \foreach \peak in {163, 122}
    {
        \foreach \bins/\loss/\losss in {3/145/243,4/103/276,5/32/285,6/19/267}
        {
            \begin{tikzpicture}
                \begin{axis} [
                    width=0.33\textwidth,
                    ymode = normal,
                    xticklabels={,,},
                    yticklabels={,,},
                    enlargelimits=false,
                    axis equal image,
                    axis on top,
                    axis line style={draw=none},
                    title=\ifthenelse{\equal{\peak}{163}}{{$\log p(N_b|\mathcal{D})=\loss$}}{$\log p(N_b|\mathcal{D})=\losss$},
                    tick style={draw=none},
                    very thick ]
                    \addplot graphics [points={(0,0) (1,1)}, includegraphics={trim=0 0 0 0,clip}] {{images/weak_id_\peak_\bins.png}};
                \end{axis}
            \end{tikzpicture}
        }
        \\[0.1em]
    }
    \caption{Optimal binning for weak (top) and strong (bottom) reflections using the Knuth's algorithm in \eqref{eq:bins_posterior}.}
    \label{fig:knuth_bins}
\end{figure*}
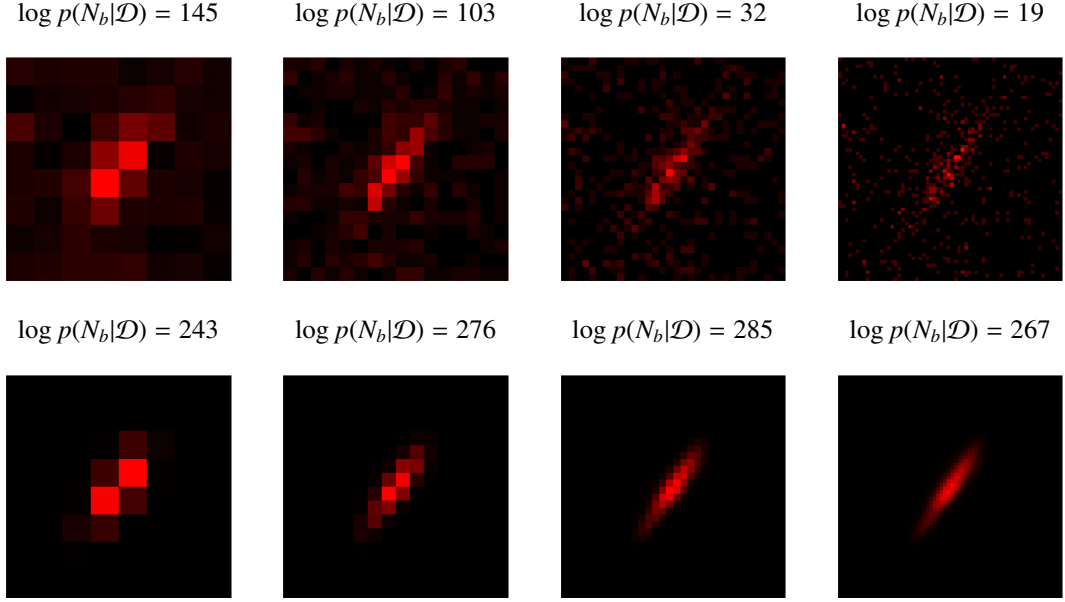

\subsection{Coarsest scale}

The choice of the coarsest scale for $\mathcal{H}_0$ determines the level of details that can be resolved accurately.
Stronger reflections contain sufficient amount of fine scale information to be resolved at higher resolutions.
In this case, the choice of inappropriate $\mathcal{H}_0$ can degrade the quality of fit.
On the contrary, weak reflections can be properly resolved only at coarser resolutions requiring a corresponding choice of $\mathcal{H}_0$.
To resolve this issue in a consistent way for both strong and weak reflections, we consider the problem of optimal histogram binning.
This question has been approached in the literature from different perspectives and usually involves solving an optimization problem with notable examples given by the classical asymptotic rules that are optimal for normal distributions \cite{enwiki:1215210750,SCOTT1979,Wand1997}, and more recent data-driven approaches that work better for arbitrary distributions \cite{Scargle_1998,Shimazaki2007,galbrun2022minimum}.
Here we propose to use a variant of the Knuth's method that is well-suited for higher dimensional histograms with uniform bins in each dimension \cite{KNUTH2019102581}.
The method considers a multinomial likelihood with Jeffrey’s prior for the bin counts and solves for the optimal number of bins which gives
\begin{align*}
    \hat{N}_b = \argmax_{N_b} \log p(N_b|\mathcal{D}) .
\end{align*}
with
\begin{align}\label{eq:bins_posterior}
    &p(N_b|\mathcal{D}) 
    \propto \frac{N_b^{d\cdot N_e}\Gamma\left(\frac{N_b^d}{2}\right)}{\Gamma\left(\frac{1}{2}\right)^{N_b^d} \Gamma\left(N_e+\frac{N_b^d}{2}\right)}
    \prod_{i=1}^{N_b^d} \Gamma\left(n_{i}+\frac{1}{2}\right)
\end{align}
assuming equal number of bins $N_b$ in each out of $d$ dimensions; the product is over an arbitrary enumeration of $d$-dimensional bins.
Figure~\ref{fig:knuth_bins} illustrates the optimal scale for selection weak and strong reflections using the Knuth's algorithm.

\begin{algorithm*}[tb]
\caption{Hierarchical profile fitting}\label{alg:fitting_pseudocode}
\begin{algorithmic}
    \State \textbf{Given:} box size with a single peak, prior $p(\lambda)$, iterative optimizer, regularization weight $\alpha\geq 1$
    \State $\hat{N}_b \gets \argmax_{N_b} \log p(N_b|\mathcal{D})$\Comment{Find the optimal binning of $\mathcal{H}_0$ using \eqref{eq:optimal_bins}}
    \State $\lambda_{0}^0 \gets p(\lambda)$\Comment{Initialize parameters}
    \State $\lambda_0 \gets \argmax_{\lambda} \log\mathcal{L}(\mathcal{H}_0|\lambda) + \log p(\lambda)$\Comment{Fit at coarsest scale $\mathcal{H}_0$}
    \For{$s=1,\hdots,L$}\Comment{Hierarchical refinement}
        \State $\lambda_s^0 \gets \lambda_{s-1}$ \Comment{Reinitialize parameters using previous scale}
        \State $\lambda_s \gets \argmax_{\lambda} \log\mathcal{L}(\mathcal{H}_s|\lambda) + \alpha\cdot\log p(\lambda|\mathcal{H}_0,\hdots,\mathcal{H}_{s-1})$\Comment{Fit at current scale $\mathcal{H}_s$}
    \EndFor
\end{algorithmic}
\end{algorithm*}

\section{Algorithm}
\label{sec:alg}

The high level summary of the proposed profile fitting approach is given in Algorithm~\ref{alg:fitting_pseudocode}. 
Next sections provide implementation details of each step.

\subsection{Rate parameterization}

We choose ellipsoidal representation of the peak shape with uniform background.
This gives
\begin{align*}
    \lambda(q) = b^2 + s^2 \exp\left( -\frac{1}{2} \left\| \sqrt{P} \cdot (q-\mu) \right\|^2 \right),
\end{align*}
where $b^2$ is the positive constant background, $s^2$ is the maximum peak intensity, $\mu$ is the location of the peak center, and $\sqrt{P}$ is the matrix encoding the shape of the peak such that $P=\sqrt{P}^T\sqrt{P}$ is the precision matrix of the Gaussian.
The total number of parameters for each peak is $11$.
The precision matrix of the peak ellipsoid is parameterized using Givens rotations as follows \cite{pinheiro1996unconstrained}
\begin{align*}
    \sqrt{P} = \sqrt{D^{-1}} \cdot R
    \quad\to\quad
    P = R^T \cdot D^{-1} \cdot R,
\end{align*}
where $R$ is the unitary rotation matrix
\begin{align*}
    &R(\varphi_1,\varphi_2,\varphi_3) = R(\varphi_3) \cdot R(\varphi_2) \cdot R(\varphi_1),
    \\
    &R(\varphi_1) =
    \begin{bmatrix}
    1 &  0   &  0   \\
    0 &  c_1 & -s_1 \\
    0 &  s_1 &  c_1
    \end{bmatrix},
    \quad
    R(\varphi_2) =
    \begin{bmatrix}
     c_2 & 0 & -s_2\\
     0   & 1 &  0  \\
     s_2 & 0 &  c_2
    \end{bmatrix},
    \\
    &R(\varphi_3) =
    \begin{bmatrix}
     c_3 & -s_3 & 0\\
     s_3 &  c_3 & 0\\
     0   &  0   & 1 
    \end{bmatrix},
\end{align*}
and $$\sqrt{D}=diag(\sigma_1,\sigma_2,\sigma_3)$$ is the diagonal matrix with the semi-axes of the ellipsoid.
This choice of parametric model for the flux $\lambda$ is easily interpretable and expressive but simple enough so that the optimization problem can be solved efficiently and robustly. 

Once the optimal parameters of the Gaussian fit are estimated from the data, the peak region is identified with the points $q\in\mathbf{Q}$ satisfying
\begin{align*}
    \left\| \sqrt{P} \cdot (q-\mu) \right\| \leq \sigma_{peak},
\end{align*}
and the background is estimated from the ellipsoidal shell 
\begin{align*}
    \sigma_{peak} < \left\| \sqrt{P} \cdot (q-\mu) \right\| \leq \sigma_{bkgr},
\end{align*}
where $\sigma_{peak},\sigma_{bkgr}$ are the standard deviations of the Gaussian ellipsoids enclosing the peak and background regions.
The natural choice for the peak region bound is $\sigma_{peak}=4$ to ensure that all relevant data points are accounted for.
The upper bound for the background ellipsoidal shell can be varied, but we found that $\sigma_{bkgr}=3\sigma_{peak}$ is the adequate choice.

\subsection{Optimization problem}

The prior $p(\lambda)$ encodes any available knowledge about the peak shape prior to actual fitting.
An example of information that is usually known is given by the physical constraints on the possible size of the peak.
Given this information, we arrive at the constrained optimization problem with box constraints for the intensity $\lambda(\mathbf{p})$ with parameters $p_i,\;i=1,\hdots,11,$
\begin{align*}
    \max_{\mathbf{p}} \mathcal{L}(\mathcal{H}_s|\lambda) + p(\lambda|\mathcal{H}_{s-1})
    \quad\text{s.t.}\quad \mathbf{p}:=\{p_i:l_i\leq p_i\leq u_i\},
\end{align*}
where not all bounds are necessarily given.

\subsection{Integration box}

The sizes of Bragg reflections within the same dataset can vary substantially but always have a hard upper bound imposed by physical constraints.
The choice of the integration box must respect these constraints but also should account for the peak size variation in order to keep the ratio of the peak vs background data comparable for all peaks.

Given the ellipsoidal parameterizarion of the $j$-th peak shape, we select the integration box with the size
\begin{align*}
    |box|_j = \min \left( |box|_{max}, 2\|\mu_j-\mu_{nn}\|, \sigma_{bkgr}\cdot\max_i\sqrt{D_{ii}} \right)    
\end{align*}
to ensure the box is large enough to contain both the peak and the background shell of exactly one peak.
The distance $\|\mu_j-\mu_{nn}\|$ measures the distance from the current peak to the center of the closest peak.

\subsection{Detector mask}

TOPAZ detector arrays do not provide full coverage of the $\mathbf{Q}$-space.
This means that not all the points in the integration box correspond to the actual region on the detector plates.
Moreover, the binned data will produce zero counts for the voxels outside the detector.
Without explicitly masking out such regions, both the fit and integration results will be biased towards zero hindering reliability of the proposed technique.
Fortunately, every location in $\mathbf{Q}$-space can be checked to be inside or outside of the detector region.
However, this requires solving an expensive ray-tracing problem with cubic complexity in terms of the histogram size which quickly becomes the computational bottleneck.

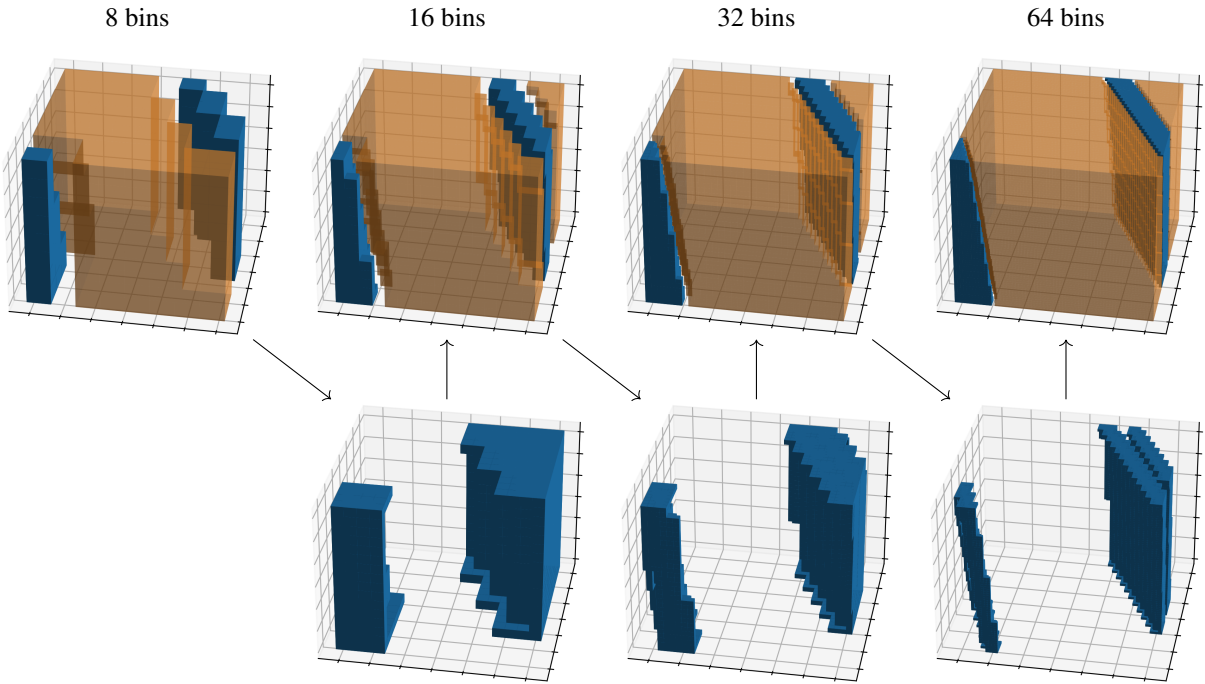
\begin{figure*}[tb]
    \centering
    \foreach \name in {nodetector,boundary}
    {
        \foreach \peak in {227}
        {
            \foreach \bins in {8,16,32,64}
            {
            {
                \begin{tikzpicture}[trim axis left,trim axis right]
                    \begin{axis} [
                        width=0.38\textwidth,
                        ymode = normal,
                        xticklabels={,,},
                        yticklabels={,,},
                        enlargelimits=false,
                        axis equal image,
                        axis on top,
                        axis line style={draw=none},
                        title=\ifthenelse{\equal{\name}{nodetector}}{{\bins} bins}{},
                        tick style={draw=none},
                        very thick ]
                        \IfFileExists{{images/peak_\peak_bins_128_\name_mask_\bins.png}}{
                        \addplot graphics [points={(0,0) (1,1)}, includegraphics={trim=30 30 30 30,clip}] {{images/peak_\peak_bins_128_\name_mask_\bins.png}};
                        }{}
                        \node[] at (0.9,0.0) {\tikzmark{tail\name\bins}};
                        \node[] at (0.1,1.0) {\tikzmark{head\name\bins}};

                        \node[] at (0.5,0.0) {\tikzmark{head2\name\bins}};
                        \node[] at (0.5,1.0) {\tikzmark{tail2\name\bins}};
                    \end{axis}
                \end{tikzpicture}
            }}
        }
        \\[0.1em]
    }
    \vspace{-0.1em}
    \begin{tikzpicture}[remember picture,overlay]
        \draw[->] (pic cs:tailnodetector8)  -- (pic cs:headboundary16);
        \draw[->] (pic cs:tailnodetector16) -- (pic cs:headboundary32);
        \draw[->] (pic cs:tailnodetector32) -- (pic cs:headboundary64);

        \draw[->] (pic cs:tail2boundary16)  -- (pic cs:head2nodetector16);
        \draw[->] (pic cs:tail2boundary32)  -- (pic cs:head2nodetector32);
        \draw[->] (pic cs:tail2boundary64)  -- (pic cs:head2nodetector64);
    \end{tikzpicture}
    \caption{Hierarchical probing of the detector mask. Transparent and blue voxels are inside and outside of the detector region respectively.}
    \label{fig:detector_mask}
\end{figure*}

Instead, we employ a hierarchical probing technique summarized in Figure~\ref{fig:detector_mask}.
The method starts with a very coarse histogram, e.g., of size $8^3$ bins, and checks every voxel with a ray tracing approach.
Given this estimate, the boundary of the detector region is determined as the voxels that have neighbors both inside and outside of the detector, or at the integration box boundary.
At the next step, the higher resolution histogram is generated and only the boundary voxels from the previous iteration are checked.
The method proceeds iteratively until the required resolution is achieved.
The method also fits perfectly into the proposed hierarchical integration framework since the detector mask needs to be refined only when the relevant resolution has been reached.

\subsection{Initialization}

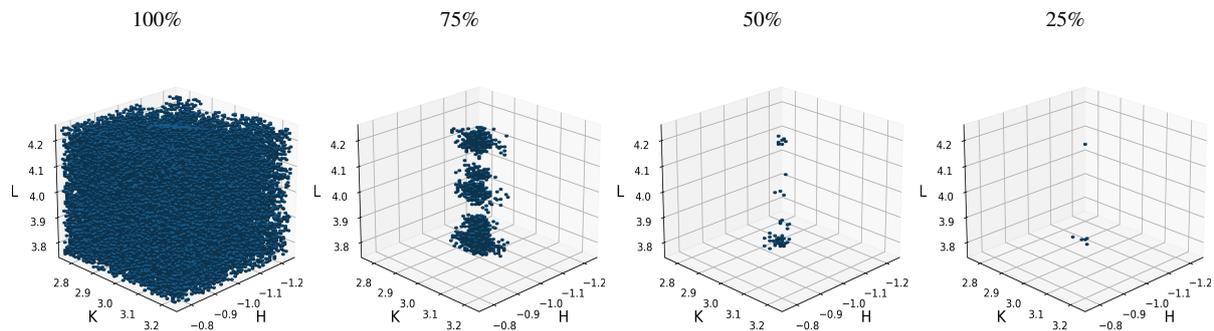
\begin{figure*}[tb]
    \pgfplotsset{every tick label/.append style={font=\scriptsize}}
    \pgfplotsset{every axis label/.append style={font=\footnotesize}}
    \pgfplotsset{every axis title/.append style={font=\footnotesize}} 
    \centering
    \foreach \peak in {62}
    {
        \foreach \step/\ratio in {20/100,16/75,10/50,5/25}
        {
            \begin{tikzpicture}[trim axis left,trim axis right]
                \begin{axis} [
                    width=0.38\textwidth,
                    ymode = normal,
                    xticklabels={,,},
                    yticklabels={,,},
                    enlargelimits=false,
                    axis equal image,
                    axis on top,
                    axis line style={draw=none},
                    title={$\ratio \%$},
                    tick style={draw=none},
                    very thick ]
                    \IfFileExists{{images/peak_\peak_bins_64_bins_64_filtered_mask_\step.png}}{
                    \addplot graphics [points={(0,0) (1,1)}, includegraphics={trim=9 9 9 9,clip}] {{images/peak_\peak_bins_64_bins_64_filtered_mask_\step.png}};
                    }{}
                \end{axis}
            \end{tikzpicture}
        }
    }
    \caption{Filtration of the histogram by intensity values.}
    \label{fig:intensity_filtration}
\end{figure*} 

\begin{algorithm}[tb]
\caption{Estimate enclosing sphere}\label{alg:sphere}
\label{alg:enclosing_sphere}
\begin{algorithmic}
    \State \textbf{Given:} histogram $\mathcal{H}_0$, peak center $\mu$, number of threshold radii $n$ 
    \State $\{a, b\} \gets \{\displaystyle\min_x \mathcal{H}_0[x], \displaystyle\max_x \mathcal{H}_0[x]\}$\Comment{Threshold bounds}
    \For{$i\gets 0,n$} 
        \State $t_i \gets a + i(b-a)/n$\Comment{Calculate threshold value}
        \State $r_i \gets \displaystyle\max_{\mathcal{H}_0[x]\geq t_i} \|x-\mu\|$\Comment{Largest enclosing sphere}
    \EndFor
    \State $i^* \gets \displaystyle\argmax_{i\in [1,n-1]} |r_i-r_{i-1}|$\Comment{Largest drop in radius}
    \State $r^* \gets r_{i^*}$\Comment{Estimated radius of enclosing sphere}
\end{algorithmic}
\end{algorithm}

To initialize the parameters of the model, we employ a simple and computationally cheap heuristic.
It starts with the intensity filtration of the histogram $\mathcal{H}_0$ to estimate the voxels with intensities above the uniform background, see Algorithm~\ref{alg:enclosing_sphere}.
Given positions $x_i$ of such voxels, we approximately solve the minimum volume enclosing ellipsoid (MVEE) problem given by
\begin{align*}
    \text{minimize } &\quad-\log\det C
    \\
    \text{subject to } &\quad x_i^TCx_i\leq 1,\;\forall i,
\end{align*}
where $C=P^{-1}$ is the covariance matrix of the ellipsoid.
Other algorithms for estimating ellipsoids can be also used, but MVEE is well suited for the relatively small number of voxels at level $\mathcal{H}_0$.
It also doesn't tend to underestimate the size of the ellipsoid since the method is non-statistical and no voxels are assumed to stay outside the detected ellipsoid.
Also, we solve MVEE only approximately using a small ($<10$) number of iterations of an appropriate solver \cite{bowman2023computing}.

Given the covariance matrix $C=P^{-1}$, the background intensity $b^2$ is initialized to the mean of $\mathcal{H}_0$ in the region outside the $\sigma_{peak}$ ellipsoid. This estimate is in fact an MLE of the uniform background model
\begin{align*}
    \lambda(x) = b^2
    \quad\to\quad
    \log\mathcal{L}(\mathcal{H}_0|b^2) = \sum_{i} n_i \log b^2 - b^2 
\end{align*}
which attains its minimum at the mean intensity.
Given the background estimate, the scale parameter $s^2$ is set to the background corrected intensity of the histogram at the center of the ellipsoid.

\begin{figure*}[htb]
    \pgfplotsset{every tick label/.append style={font=\scriptsize}}
    \pgfplotsset{every axis label/.append style={font=\footnotesize}}
    \pgfplotsset{every axis title/.append style={font=\footnotesize}} 
    \centering
    \begin{subfigure}{\textwidth}
        \foreach \peak in {62}
        {
            \foreach \step in {1,2,3,4}
            {
                \begin{tikzpicture}[trim axis left,trim axis right]
                    \begin{axis} [
                        width=0.38\textwidth,
                        ymode = normal,
                        xticklabels={,,},
                        yticklabels={,,},
                        enlargelimits=false,
                        axis equal image,
                        axis on top,
                        axis line style={draw=none},
                        tick style={draw=none},
                        very thick ]
                        \IfFileExists{{images/peak_\peak_bins_64_bins_64_shell_mask_\step.png}}{
                        \addplot graphics [points={(0,0) (1,1)}, includegraphics={trim=9 9 9 9,clip}] {{images/peak_\peak_bins_64_bins_64_shell_mask_\step.png}};
                        }{}
                    \end{axis}
                \end{tikzpicture}
            }
        }
        \caption{Shells around the peak to remove background artifacts.}
        \label{fig:artifact_shells}
    \end{subfigure}
    \begin{subfigure}{\textwidth}
        \foreach \peak in {62}
        {
            \foreach \step in {1,2,3,4}
            {
                \begin{tikzpicture}[trim axis left,trim axis right]
                    \begin{axis} [
                        width=0.38\textwidth,
                        ymode = normal,
                        xticklabels={,,},
                        yticklabels={,,},
                        enlargelimits=false,
                        axis equal image,
                        axis on top,
                        axis line style={draw=none},
                        tick style={draw=none},
                        very thick ]
                        \IfFileExists{{images/peak_\peak_bins_64_bins_64_outlier_mask_\step.png}}{
                        \addplot graphics [points={(0,0) (1,1)}, includegraphics={trim=9 9 9 9,clip}] {{images/peak_\peak_bins_64_bins_64_outlier_mask_\step.png}};
                        }{}
                    \end{axis}
                \end{tikzpicture}
            }
        }
        \caption{Peak with outliers detected in the shells above.}
        \label{fig:filtered_shells}
    \end{subfigure}
    \caption{Relatively weak peak with strong background artifacts.}
    \label{fig:weak_peak_with_artifacts}
\end{figure*}
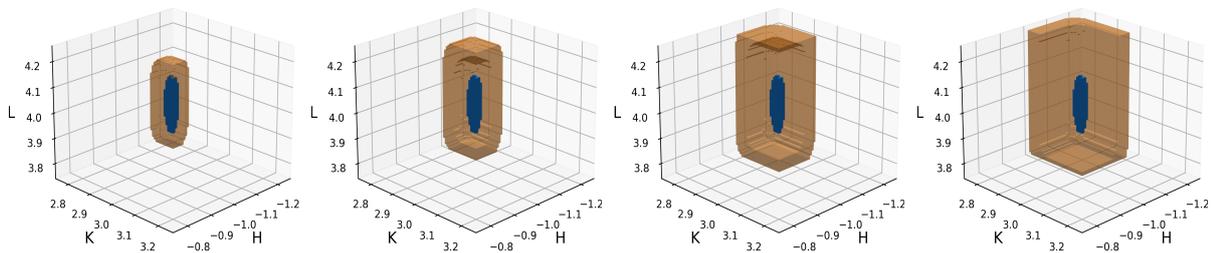
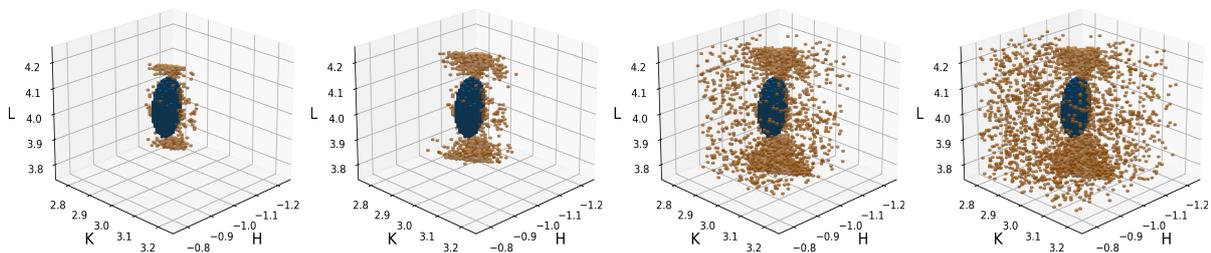

\subsection{Background filtering}

Figure~\ref{fig:intensity_filtration} illustrates an important issue that needs to be resolved before applying any of the steps above.
Specifically, the depicted histogram has a true peak in the middle and two diffuse scattering artifacts at the top and the bottom.
If these artifacts are not removed, the true peak size will be overestimated unavoidably corrupting the integrated intensity.

We approach this problem by considering a sequence of shells around the peak region as demonstrated in Figure~\ref{fig:artifact_shells}.
In each shell, we adopt the conventional standard deviation test to mask out the voxels with intensities above three standard deviations from the mean intensity of the shell.
In each shell, we use this rule recursively with increasing number of passes, e.g., one pass for the first shell, two passes for the second shell, and so on.
Figure~\ref{fig:filtered_shells} shows the result of applying this procedure to the aforementioned peak.

The recurrent filtering in outer shells allows to penalize strong artifacts far from the peak while reducing the possibility of incorrectly removing important data from the actual peak.
This method requires an estimate of the peak shape and should be performed iteratively starting from the spherical approximation at the initialization step, followed by MVEE approximation, and iterative fits during the hierarchical optimization.
It should be noted, however, that this method must be applied only to sufficiently coarse histograms that preserve spatial correlations in data.
In practice, it is save to use this method for the first two resolution levels, i.e., $\mathcal{H}_0$ and $\mathcal{H}_1$, and use the obtained mask for higher resolution histograms.

\section{Preliminary results}
\begin{figure*}[tb]
    \centering
    \foreach \peak in {227, 122}
    {
        \foreach \bins/\loss/\losss in {3/145/243,4/103/276,5/32/285,6/19/267}
        {
            \begin{tikzpicture}
                \begin{axis} [
                    width=0.33\textwidth,
                    ymode = normal,
                    xticklabels={,,},
                    yticklabels={,,},
                    enlargelimits=false,
                    axis equal image,
                    axis on top,
                    axis line style={draw=none},
                    tick style={draw=none},
                    very thick ]
                    \addplot graphics [points={(0,0) (1,1)}, includegraphics={trim=0 0 0 0,clip}] {{images/fit_\peak_\bins.png}};
                \end{axis}
            \end{tikzpicture}
        }
        \\[0.1em]
    }
    \caption{Hierarchical fitting for weak (top) and strong (bottom) reflections.}
    \label{fig:hierarchical_fit}
\end{figure*}
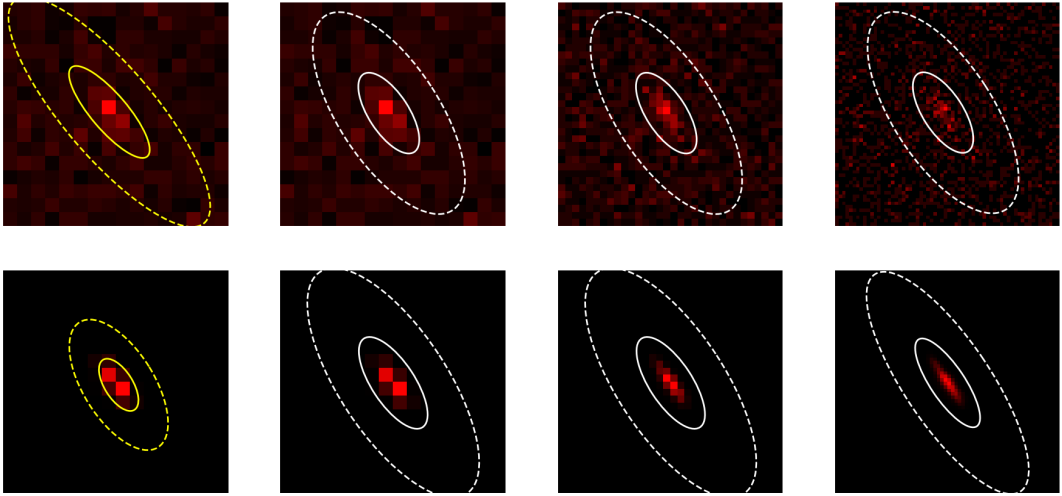

\begin{figure*}[tb]
	\centering
	\includegraphics[width=0.7\textwidth]{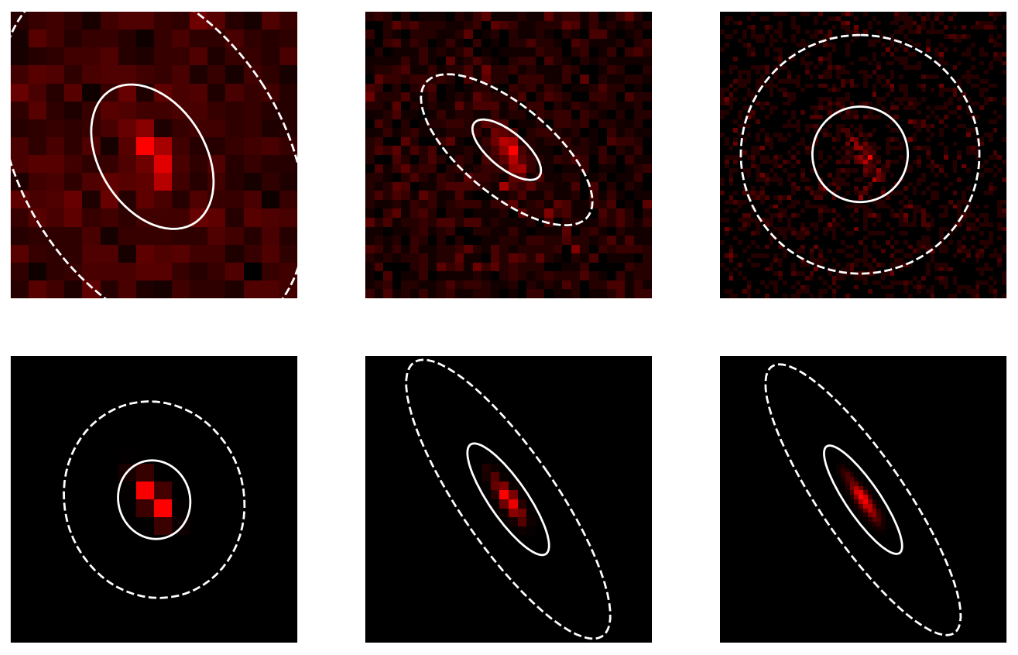}
    \caption{Direct fitting for weak (top) and strong (bottom) reflections.}
    \label{fig:direct_fit}
\end{figure*}

Figure~\ref{fig:hierarchical_fit} illustrates the result of applying the proposed method to the weak and strong peaks.
One can clearly see from the comparison with Figure~\ref{fig:direct_fit}, that hierarchical approach is consistent across the scales while the direct approach fails for weak peaks at higher resolutions.

\bibliographystyle{vancouver-authoryear}
\bibliography{biblio}





\end{document}